\documentclass[twocolumn,aps,pra,showpacs]{revtex4}

\begin{document}
\title{Six-qubit permutation-based decoherence-free orthogonal basis}
\author{Ad\'{a}n Cabello}
\email{adan@us.es}
\affiliation{Departamento de F\'{\i}sica Aplicada
II, Universidad de Sevilla, 41012 Sevilla, Spain}
\date{\today}


\begin{abstract}
There is a natural orthogonal basis of the 6-qubit decoherence-free
(DF) space robust against collective noise. Interestingly, most of
the basis states can be obtained from one another just permuting
qubits. This property: (a) is useful for encoding qubits in DF
subspaces, (b) allows the implementation of the Bennett-Brassard
1984 (BB84) protocol in DF subspaces just permuting qubits, which
completes a the method for quantum key distribution using DF states
proposed by Boileau {\em et al.} [Phys. Rev. Lett. {\bf 92}, 017901
(2004)], and (c) points out that there is only one 6-qubit DF state
which is essentially new (not obtained by permutations) and
therefore constitutes an interesting experimental challenge.
\end{abstract}



\maketitle

\date{\today}


\section{Introduction}


Quantum communication and computation~\cite{BD00,NC00} is based on
the preparation and manipulation of qubit states. However, qubit
states are very fragile and easily destroyed by decoherence due to
unwanted coupling with the environment~\cite{Zurek91}. There are
several strategies to deal with decoherence, each of them
appropriate for a specific type of coupling with the environment. On
the one hand, if the interaction with the environment is weak enough
so there is only a low probability of the qubits being affected,
then a good strategy is to add redundancy when encoding the quantum
information and correct the errors by using active quantum error
correction methods~\cite{Shor95,Steane96,BDSW96,LMPZ96}. On the
other hand, not all states are equally fragile when interacting with
the environment. Indeed, if the qubit-environment interaction
exhibits some symmetry, there are states which are immune to this
interaction and can therefore be used to protect quantum
information. These states are called decoherence-free (DF) states
\cite{PSE96,ZR97a,LCW98,LBKW00,KBLW01}.

A particularly relevant symmetry arises in the so-called {\em
collective} noise, where the environment couples with the qubits
without distinguishing between them. This situation occurs naturally
when the spatial (temporal) separation between the qubits is small
relative to the correlation length (time) of the environment.
Typical examples arise in ion-trap or nuclear magnetic resonance
(NMR) experiments suffering fluctuations of magnetic or
electrostatic fields and also when polarized photons are
successively sent via the same optical fiber which, due to thermal
or stress variations, introduces an uncontrollable (but the same)
birefringence.


DF states immune to collective noise (hereafter simply referred as
DF states) are invariant under any $n$-lateral unitary
transformation (i.e., $ U^{\otimes n} |\psi\rangle= |\psi\rangle$,
where $U^{\otimes n} = U \otimes \ldots \otimes U$ denotes the
tensor product of $n$ unitary transformations $U$) \cite{ZR97a}.
This property makes them also useful for quantum information
processing between parties who do not share a common reference
frame. Specifically, they can be used for quantum key distribution
\cite{BGLPS04,CZBJYZYLP06} and for other communication protocols
involving two \cite{BRS03,Cabello03a} or more \cite{Cabello03b}
parties who do not share any reference frame.


The amount of quantum information that a given DF subspace is able
to protect depends on the number $N$ of qubits. For $N$ {\em even},
the DF subspace spanned by states which are eigenstates of the whole
Hamiltonian of the qubits-bath system and also eigenstates of the
interaction Hamiltonian with eigenvalue zero has dimension
\cite{ZR97a}
\begin{equation}
d(N)={N! \over (N/2)! (N/2+1)!}.
\label{dim}
\end{equation}
The number of qubits encoded in DF states is $\log_2 d(N)$. For a
large $N$,
\begin{equation}
\log_2 d(N) \simeq N-{3 \over 2} \log_2 N.
\end{equation}
Therefore, the encoding efficiency is asymptotically unity.


For $N=2$ qubits, there is only one DF state, the singlet state
\begin{equation}
|\psi^-\rangle = \frac{1}{\sqrt{2}} (|01\rangle-|10\rangle),
\end{equation}
where $|01\rangle=|0\rangle \otimes |1\rangle$. Several experiments
have demonstrated the invariance properties of the singlet and its
immunity against $U \otimes U$
\cite{KBAW00,KMRSIMW01,FVHTC02,OLK03}.

For $N=4$ qubits, the dimension of the DF subspace is $2$.
Therefore, $N=4$ qubits are sufficient to fully protect one
arbitrary logical qubit against collective noise. A natural choice
of orthogonal basis is the one containing the double singlet,
denoted by $|\psi^-\rangle_{12}\otimes |\psi^-\rangle_{34}$, and the
only DF state which is orthogonal to it. This state can be
calculated by applying the Gram-Schmidt orthogonalization method to
the double singlet and any other state invariant under $U^{\otimes
4}$---for instance, the one obtained from the double singlet
permuting qubits $2$ and $3$, denoted by $|\psi^-\rangle_{13}\otimes
|\psi^-\rangle_{24}$. The resulting state turns out to be the
4-qubit ``supersinglet'' \cite{Cabello03}. This leads to the
following orthogonal basis of the $4$-qubit DF subspace:
\begin{eqnarray}
|\bar{0}\rangle & \equiv & |\psi^-\rangle_{12}\otimes
|\psi^-\rangle_{34},\\
|\bar{1}\rangle & \equiv & {1 \over 2 \sqrt{3}}
(2|0011\rangle-|0101\rangle-|0110\rangle \nonumber \\
& & -|1001\rangle -|1010\rangle+2|1100\rangle).
\end{eqnarray}
This basis was first proposed in \cite{KBLW01}. Preparing the double
singlet just requires the duplication of the setup to prepare a
singlet state. Preparing the new state $|\bar{1}\rangle$ was an
interesting challenge. Bourennane {\em et al.} did it by spontaneous
parametric down-conversion and, using a different setup, they also
prepared the double singlet \cite{BEGKCW04}. They demonstrated the
immunity of both states against collective noise by showing their
invariance when passing the four photons through a noisy environment
simulated by birefringent media. In addition, they showed that these
two orthogonal DF states can be distinguished by fixed (i.e., not
conditioned \cite{Cabello03b}) one-qubit polarization measurements.
However, since each state requires a different setup, a still open
problem is that of encoding an arbitrary logical qubit in a
polarization DF subspace.


\section{Six-qubit decoherence-free basis}


\subsection{Decoherence-free subspace spanned by
permutations of 2- and 4-qubit states}


Some experimental groups are developing sources of {\em six}-photon
polarization-entangled states \cite{ZGWCZYMSP06,LZGGZYGYP07}. One
remarkable point is that $N=6$ qubits make room for a DF subspace of
dimension $5$ [see Eq. (\ref{dim})]. This number is interesting for
two reasons: first, because it is the minimum needed to fully
protect {\em two} arbitrary logical qubits against collective noise
and specifically to fully protect any arbitrary two-qubit {\em
entangled} state and, second, because the extra dimensions of the DF
subspace can be useful for encoding arbitrary qubits in the DF
subspace. This hope is based on the fact that there is a very
economical (in terms of the number of required experimental setups)
way to generate 6-qubit DF orthogonal states. The $N=6$ case is the
first one in which orthogonal DF states can be obtained just {\em
permuting} qubits. Indeed, one can prepare up to {\em four} mutually
orthogonal 6-qubit DF states just by combining the two setups of
Ref.~\cite{BEGKCW04}. A natural choice for these basis states is the
following:
\begin{eqnarray}
|\bar{0}\bar{0}\bar{0}\rangle & \equiv & |\psi^-\rangle_{12} \otimes
|\psi^-\rangle_{34} \otimes |\psi^-\rangle_{56},
\label{DF000}\\
|\bar{0}\bar{1}\bar{1}\rangle & \equiv & |\psi^-\rangle_{12} \otimes
|\bar{1}\rangle_{3456},
\label{DF011}\\
|\bar{1}\bar{0}\bar{1}\rangle & \equiv & |\psi^-\rangle_{34} \otimes
|\bar{1}\rangle_{1256},
\label{DF010}\\
|\bar{1}\bar{1}\bar{0}\rangle & \equiv & |\psi^-\rangle_{56} \otimes
|\bar{1}\rangle_{1234}, \label{DF110}
\end{eqnarray}
where the subindices express that some states are obtained from a
single one permuting qubits: for instance,
\begin{eqnarray}
|\bar{1}\bar{0}\bar{1}\rangle & = & P_{24}
P_{13}|\bar{0}\bar{1}\bar{1}\rangle, \\
|\bar{1}\bar{1}\bar{0}\rangle & = & P_{26}
P_{15}|\bar{0}\bar{1}\bar{1}\rangle,
\end{eqnarray}
where $P_{ij}$ means permuting qubits $i$ and $j$. This possibility
leads to two observations. The first is that it seems feasible to
encode arbitrary qubits exploiting the fact that there are (three)
orthogonal DF states which are obtained from one another permuting
$4$ qubits (i.e., by two permutations). Specifically, designing a
setup capable of preparing, for instance, states like
\begin{equation}
|\Psi\rangle = (\cos \theta + e^{i \phi} \sin \theta P_{24} P_{13})
|\bar{0}\bar{1}\bar{1}\rangle,
\end{equation}
by making some paths indistinguishable and appropriately combining
them would be an interesting experimental challenge.


\subsection{Genuine 6-qubit decoherence-free state}


The other observation is that the dimension of the DF subspace which
is {\em not} spanned by the four states (\ref{DF000})--(\ref{DF110})
is $1$, meaning that there is just one DF state that cannot be
prepared by combining previous setups. This state can be calculated
using the Gram-Schmidt method. The missing state is
\begin{eqnarray}
|\bar{1}\bar{1}\bar{1}\rangle & \equiv & {1 \over 2\sqrt {3}}
\sum_{\scriptscriptstyle{ {\stackrel{\scriptscriptstyle{\rm
permutations}} {{\rm of}\;000111}}}} \!\!\!\!\!\!(-1)^t\left|ijklmn
\right\rangle, \label{DF111}
\end{eqnarray}
where $t$ is the number of transpositions of pairs of elements that
must be composed to place the elements in canonical order (i.e.,
$000111$). Therefore, another interesting challenge would be to
describe a setup for preparing this genuinely new $6$-qubit DF
state.

Another interesting property of the orthogonal basis of the
$6$-qubit DF subspace composed by the states
(\ref{DF000})--(\ref{DF110}) and (\ref{DF111}) is that it is
possible to distinguish any two basis states by fixed single qubit
measurements. For instance, $|\bar{0}\bar{1}\bar{1}\rangle$ and
$|\bar{1}\bar{0}\bar{1}\rangle$ can be distinguished by measuring
$\sigma_z \otimes \sigma_z \otimes \sigma_x \otimes \sigma_x \otimes
\sigma_z \otimes \sigma_z$. The single-qubit measurements that allow
us to distinguish any two basis states are summarized in Table I.


\begin{table}[tbp]
\begin{center}
\begin{ruledtabular}
\begin{tabular}{cccccc}
& $|\bar{0}\bar{0}\bar{0}\rangle$ & $|\bar{0}\bar{1}\bar{1}\rangle$
& $|\bar{1}\bar{0}\bar{1}\rangle$ & $|\bar{1}\bar{1}\bar{0}\rangle$
& $|\bar{1}\bar{1}\bar{1}\rangle$ \\
\hline $|\bar{0}\bar{0}\bar{0}\rangle$ & &
$zzxxzz$ & $zzzzxx$ & $xxzzzz$ & $zzzzzz$ \\
$|\bar{0}\bar{1}\bar{1}\rangle$ & $zzxxzz$ & & $zzxxzz$ &
$zzzzxx$ & $xxzzzz$ \\
$|\bar{1}\bar{0}\bar{1}\rangle$ & $zzzzxx$ & $zzxxzz$ & & $zzxxzz$ &
$zzxxzz$ \\
$|\bar{1}\bar{1}\bar{0}\rangle$ & $xxzzzz$ & $zzzzxx$ & $zzxxzz$ & &
$zzzzxx$ \\
$|\bar{1}\bar{1}\bar{1}\rangle$ & $zzzzzz$ & $xxzzzz$ & $zzxxzz$ &
$zzzzxx$ & \\
\end{tabular}
\end{ruledtabular}
\end{center}
\noindent TABLE I. {\small Measurements that allow us to distinguish
any pair of the 6-qubit DF states defined in Eqs.
(\ref{DF000})--(\ref{DF110}) and (\ref{DF111}). For instance,
$zzxxzz$ means that the two states in the corresponding column and
row can be distinguished by measuring $\sigma_z \otimes \sigma_z
\otimes \sigma_x \otimes \sigma_x \otimes \sigma_z \otimes
\sigma_z$.}
\end{table}


\subsection{BB84 protocol using permutations of a single 6-qubit decoherence-free state}


Finally, another interesting observation is that all four states
needed for a DF version of the Bennett-Brassard 1984 (BB84) protocol
(or four-state scheme) \cite{BB84} can be obtained {\em by
permutations of a single DF state}. For instance, we can define
\begin{eqnarray}
|\hat{0}\rangle & \equiv & |\bar{0}\bar{1}\bar{1}\rangle,\\
|\hat{\oplus}\rangle & \equiv & P_{13} |\hat{0}\rangle,\\
|\hat{1}\rangle & \equiv & P_{24} |\hat{\oplus}\rangle,\\
|\hat{\ominus}\rangle & \equiv & P_{13} |\hat{1}\rangle.
\end{eqnarray}
These four states satisfy
\begin{equation}
\left|\langle \hat{\oplus}|\hat{0}\rangle\right|^2 =
\left|\langle \hat{\ominus}|\hat{0}\rangle\right|^2 =
\left|\langle \hat{\oplus}|\hat{1}\rangle\right|^2 =
\left|\langle \hat{\ominus}|\hat{1}\rangle\right|^2 = 1/2,
\end{equation}
as required for the BB84 protocol. Since both the computational
basis $\left\{|\hat{0}\rangle, |\hat{1}\rangle\right\}$ and the
Hadamard basis $\left\{|\hat{\oplus}\rangle,
|\hat{\ominus}\rangle\right\}$ can be obtained permuting qubits on a
single DF state, a setup for preparing the state
$|\bar{0}\bar{1}\bar{1}\rangle$ and a mechanism to permutate the
outputs \cite{JYCKB04} in Alice's side, and a setup for measuring
$\sigma_z \otimes \sigma_z \otimes \sigma_x \otimes \sigma_x \otimes
\sigma_z \otimes \sigma_z$ (to distinguish $|\hat{0}\rangle$ and
$|\hat{1}\rangle$) or, alternatively, one for measuring $\sigma_x
\otimes \sigma_z \otimes \sigma_z \otimes \sigma_x \otimes \sigma_z
\otimes \sigma_z$ (to distinguish $|\hat{\oplus}\rangle$ and
$|\hat{\ominus}\rangle$) in Bob's side are sufficient to implement
an exact replica of the BB84 protocol using DF states. This DF
version of the BB84 protocol completes the quantum key distribution
protocol proposed by Boileau {\em et al.} \cite{BGLPS04} which is
essentially a permutation-based DF version of the Bennett 1992 (B92)
protocol using nonorthogonal states \cite{Bennett92}. The
characteristic features of the BB84 protocol derive from the fact
that it uses two mutually unbiased orthogonal bases. Two orthogonal
bases are mutually unbiased if any basis states $|e_j \rangle$ and
$|e_\mu \rangle$ belonging to different bases satisfy $\left|\langle
e_\mu | e_j \rangle\right|^2=1/2$. Therefore, each state in one of
these bases is an equal-magnitude superposition of all the states in
any of the other bases. As a consequence, if an eavesdropper (Eve)
uses an intercept-and-resend strategy and measures in the wrong
basis, she gets no information at all and causes maximal disturbance
(error rate $1/2$) to the transmission, thereby revealing her
presence \cite{BB84,BP00,ABB01,BM02}.


\section{Eight-qubit decoherence-free basis}


\subsection{Decoherence-free subspace spanned by permutations
of 2-, 4-, and 6-qubit states}


The next question is whether or not the process of generating an
orthogonal DF basis by using products of DF states in lower
dimensions and permuting qubits can be extending to higher
dimensions. The next natural step is to study the DF subspace of
$N=8$ qubits which is of dimension $14$ [see Eq. (\ref{dim})]. How
many mutually orthogonal DF states can be obtaining by combining the
states of the $6$-qubit DF subspace and permuting qubits? The answer
is that we can obtain up to $12$ orthogonal DF states by products of
lower dimensional DF states and permutations of qubits. A natural
choice of basis states is the following:
\begin{eqnarray}
|\bar{0}\bar{0}\bar{0}\bar{0}\rangle & \equiv & |\psi^-\rangle_{12}
\otimes |\psi^-\rangle_{34} \otimes |\psi^-\rangle_{56} \otimes
|\psi^-\rangle_{78},
\label{DF0000}\\
|\bar{0}\bar{0}\bar{1}\bar{1}\rangle & \equiv & |\psi^-\rangle_{12}
\otimes |\psi^-\rangle_{34} \otimes
|\bar{1}\rangle_{5678}, \\
|\bar{0}\bar{1}\bar{0}\bar{1}\rangle & \equiv & |\psi^-\rangle_{12}
\otimes |\psi^-\rangle_{56} \otimes
|\bar{1}\rangle_{3478}, \\
|\bar{0}\bar{1}\bar{1}\bar{0}\rangle & \equiv & |\psi^-\rangle_{12}
\otimes |\psi^-\rangle_{78} \otimes
|\bar{1}\rangle_{3456}, \\
|\bar{1}\bar{0}\bar{0}\bar{1}\rangle & \equiv & |\psi^-\rangle_{34}
\otimes |\psi^-\rangle_{56} \otimes
|\bar{1}\rangle_{1278}, \\
|\bar{1}\bar{0}\bar{1}\bar{0}\rangle & \equiv & |\psi^-\rangle_{34}
\otimes |\psi^-\rangle_{78} \otimes
|\bar{1}\rangle_{1256}, \\
|\bar{1}\bar{1}\bar{0}\bar{0}\rangle & \equiv & |\psi^-\rangle_{56}
\otimes |\psi^-\rangle_{78} \otimes
|\bar{1}\rangle_{1234}, \\
|\bar{1}\bar{1}\bar{1}\bar{1}\rangle & \equiv &
|\bar{1}\rangle_{1234} \otimes
|\bar{1}\rangle_{5678}, \\
|\bar{0}\bar{1}\bar{1}\bar{1}\rangle & \equiv & |\psi^-\rangle_{12}
\otimes
|\bar{1}\bar{1}\bar{1}\rangle_{345678}, \\
|\bar{1}\bar{0}\bar{1}\bar{1}\rangle & \equiv & |\psi^-\rangle_{34}
\otimes
|\bar{1}\bar{1}\bar{1}\rangle_{125678}, \\
|\bar{1}\bar{1}\bar{0}\bar{1}\rangle & \equiv & |\psi^-\rangle_{56}
\otimes
|\bar{1}\bar{1}\bar{1}\rangle_{123478}, \\
|\bar{1}\bar{1}\bar{1}\bar{0}\rangle & \equiv & |\psi^-\rangle_{78}
\otimes |\bar{1}\bar{1}\bar{1}\rangle_{123456}, \label{DF1110}
\end{eqnarray}
where the notation is the same used in Eqs.
(\ref{DF000})--(\ref{DF110}).


\subsection{Genuine 8-qubit decoherence-free states}


As in previous dimensions, there is still a DF subspace not spanned
by the states (\ref{DF0000})--(\ref{DF1110}). However, since in this
case the dimension of the subspace is $2$, there are multiple
choices for the remaining two states. However, since the 4-qubit
supersinglet was our state $|\bar{1}\rangle$, a reasonable choice
for one of the states is the 8-qubit supersinglet \cite{Cabello03}
\begin{eqnarray}
|\bar{0}\bar{0}\bar{0}\bar{1}\rangle \equiv {1 \over {4}! \sqrt
{5}}\sum_{\scriptscriptstyle{ {\stackrel{\scriptscriptstyle{\rm
permutations}} {{\rm of}\;00001111}}}} \!\!\!\!\!\! z! \left( {4}-z
\right)! (-1)^{{4}-z} \nonumber \\ \left|ijklmnpq \right\rangle,
\label{SnS2}
\end{eqnarray}
where the sum is extended to all the states obtained by permuting
the state $|00001111\rangle$ and $z$ is the number of zeros in the
first four positions. This 8-qubit supersinglet is invariant under
$U^{\bigotimes 8}$ and orthogonal to all the previous DF states
given by Eqs. (\ref{DF0000})--(\ref{DF1110}). Therefore, there is
only one additional DF state, which can be found by choosing a
suitable seed and using the Gram-Schmidt orthogonalization method.
The remaining element of the 8-qubit DF basis is
\begin{widetext}
\begin{eqnarray}
|\bar{0}\bar{0}\bar{1}\bar{0}\rangle & \equiv & \frac{1}{4 \sqrt{3}}
(|00010111\rangle + |00011011\rangle - |00011101\rangle -
|00011110\rangle + |00100111\rangle + |00101011\rangle -
|00101101\rangle - |00101110\rangle
\nonumber \\ & &
- 2|00110011\rangle + 2|00111100\rangle - |01000111\rangle -
|01001011\rangle + |01001101\rangle + |01001110\rangle +
|01110001\rangle + |01110010\rangle
\nonumber \\ & &
- |01110100\rangle - |01111000\rangle - |10000111\rangle -
|10001011\rangle + |10001101\rangle + |10001110\rangle +
|10110001\rangle + |10110010\rangle
\nonumber \\ & &
- |10110100\rangle - |10111000\rangle + 2|11000011\rangle -
2|11001100\rangle - |11010001\rangle - |11010010\rangle +
|11010100\rangle + |11011000\rangle
\nonumber \\ & &
- |11100001\rangle - |11100010\rangle + |11100100\rangle +
|11101000\rangle).
\end{eqnarray}
\end{widetext}


\section{Conclusions and further lines of research}


In conclusion, there is a natural orthogonal basis of 6-qubit DF
states with the property that almost all its elements are obtained
from a state by permuting qubits. This is interesting because: (a)
these basis states are products of previously described DF states in
lower dimensions ($2$ and $4$ qubits) that we know how to prepare
\cite{BEGKCW04}, (b) it opens the possibility of preparing arbitrary
DF qubits with a single setup, (c) the remaining DF subspace is
spanned by a single DF state, which indicates that it would be
interesting to design a setup to prepare this last state, and (d) it
allows a natural DF version of the BB84 protocol for quantum key
distribution.

In higher dimensions, no such a natural orthogonal basis exist,
since there are many possible choices. However, permutations of
lower-dimensional DF states still allow us to span most of the DF
subspace. We have proposed an almost natural basis of the next DF
subspace (i.e., the 8-qubit DF subspace) and pointed out two
orthogonal 8-qubit DF states which cannot be obtained by combination
and permutations of previous setups. For even higher dimensions, the
dimension of the DF subspace not generated by combinations and
permutations of previous states grows, so this method of generating
orthogonal basis admits multiple choices.

The main motivation of this paper has been to serve as a stimulus
for two different 6-qubit experiments: on the one hand, to stimulate the
development of an exact DF replica of the BB84 protocol based on the
preparation of a 6-qubit DF state---for instance, the state
(\ref{DF011})---and permutations of some qubits (this method
completes previous proposals for quantum key distribution using DF
states and permutations of qubits \cite{BGLPS04}) and, on the other hand,
to stimulate the preparation of the only 6-qubit DF state which
cannot be obtained by combining setups for preparing DF states in
lower dimensions and permuting qubits. This property makes the
preparation of the state (\ref{DF111}) a suitable challenge for the
recent sources of 6-photon states \cite{ZGWCZYMSP06,LZGGZYGYP07}.


\section*{Acknowledgments}


The author thanks A. Goebel and J.-W. Pan for useful conversations
and acknowledges support from Project No.~FIS2005-07689.


\end{document}